\documentclass[aps,prl,preprint,superscriptaddress,floatfix,altaffilletter]{revtex4}

\usepackage{graphicx}
\usepackage{dcolumn}
\usepackage{bm}
\usepackage{amsmath}
\usepackage{multirow}
\usepackage{hhline}
\usepackage{stackengine}

\begin{document}

\title{Stress effects on the Raman spectrum of an amorphous material: theory and experiment on a-Si:H}

\author{David A. Strubbe} \email{dstrubbe@mit.edu}
\affiliation{Department of Materials Science and Engineering, Massachusetts Institute of Technology, Cambridge, MA 02139}
\author{Eric C. Johlin} \altaffiliation{Present address: FOM Institute AMOLF, 1098 XG Amsterdam, The Netherlands}
\affiliation{Department of Materials Science and Engineering, Massachusetts Institute of Technology, Cambridge, MA 02139}
\affiliation{Department of Mechanical Engineering, Massachusetts Institute of Technology, Cambridge, MA 02139}
\author{Timothy R. Kirkpatrick}
\affiliation{Department of Mechanical Engineering, Massachusetts Institute of Technology, Cambridge, MA 02139}
\author{Tonio Buonassisi}
\affiliation{Department of Mechanical Engineering, Massachusetts Institute of Technology, Cambridge, MA 02139}
\author{Jeffrey C. Grossman} \email{jcg@mit.edu}
\affiliation{Department of Materials Science and Engineering, Massachusetts Institute of Technology, Cambridge, MA 02139}

\begin{abstract}

  Strain in a material induces shifts in vibrational frequencies, which is a probe of the nature of the vibrations and interatomic potentials,
  and can be used to map local stress/strain distributions via Raman microscopy.
  This method is standard for crystalline silicon devices, but due to lack of calibration relations, it has not been applied to amorphous materials
  such as hydrogenated amorphous silicon (a-Si:H), a widely studied material for thin-film photovoltaic and electronic devices.
  We calculated the Raman spectrum of a-Si:H \textit{ab initio} under different strains $\epsilon$
  and found peak shifts $\Delta \omega = \left( -460 \pm 10\ \mathrm{cm}^{-1} \right) {\rm Tr}\ \epsilon$. This proportionality to the trace of the strain
  is the general form for isotropic amorphous vibrational modes, as we show by symmetry analysis and explicit computation.
  We also performed Raman measurements under strain and found a consistent coefficient of $-510 \pm 120\ \mathrm{cm}^{-1}$.
  These results demonstrate that a reliable calibration for the Raman/strain relation can be achieved even for the broad peaks of an amorphous material,
  with similar accuracy and precision as for crystalline materials.

\end{abstract}

\date{\today}
\maketitle

Hydrogenated amorphous silicon (a-Si:H) is a photovoltaic material which has been studied for decades and
used commercially \cite{Jean,Shah}. Compared to the more commonly used crystalline Si (c-Si), a-Si:H has
advantages in stronger visible absorption, %
cheaper and faster fabrication, %
and the potential for flexible thin-film devices \cite{Shah}.
a-Si:H can be used alone or in heterojunction cells where it can passivate the surface of c-Si active layers \cite{Shah,George}.
It also has applications for solar water splitting \cite{Abdi}, thin-film transistors \cite{Gleskova},
bolometers \cite{Syllaios}, particle detectors \cite{Franco}, and microelectromechanical systems \cite{Chang}.
However, widespread adoption has been limited by two important disadvantages:
mobilities degrade under illumination via the Staebler-Wronski effect \cite{SWE}, and
efficiencies are significantly limited by low hole mobility \cite{Mattheis}.

Crystallization to c-Si is used to create higher-mobility microcrystalline Si ($\mu$c-Si) \cite{Mahan,KWu},
and could circumvent low hole mobility in a-Si:H by adding nanostructured charge-extraction channels \cite{Tabet}.
Conversion to denser c-Si induces stress, as does deposition \cite{Johlin2012}, thermal expansion,
or other processing. Stress is often large in thin films (and may be inhomogeneous \cite{Paillard}),
and is a critical parameter in a-Si:H as it affects mobilities \cite{Gleskova}, defects \cite{Johlin2014},
the Staebler-Wronski effect \cite{Stutzmann}, and mechanical failure properties \cite{Pomeroy}, and potentially transport via band-bending \cite{Feng}.

To understand the impact of stress effects on c-Si microelectronic devices, a standard
technique is Raman microscopy \cite{DeWolf,Bonera}, which yields a spatial distribution of stress in the device.
The Raman-active optical phonon modes in c-Si are shifted to higher frequency by compressive strain (and \textit{vice versa}),
with established coefficients \cite{Anastassakis1970,Anastassakis1990}
which are used to translate peak positions to local strain.
Raman microscopy is also commonly used for a-Si:H and $\mu$c-Si, generally for mapping the quality
or crystallinity of films via the position and width of the transverse optical (TO) peak \cite{Mahan} (analogous to the optical phonons of c-Si).
In contrast to the case for c-Si, for a-Si:H the relation between peak positions and strain has not been clear,
preventing detailed understanding of stress;
with accurate knowledge of the coefficient, these studies would be able to map stress too.
This property also serves as a probe of vibrations and interatomic potentials \cite{Fabian, Anastassakis1990}.
Stress effects on Raman peaks (also called ``piezo-Raman'' or ``phonon deformation potentials'') have been studied for various crystalline
semiconductors \cite{Anastassakis1991}. However there has been little work on amorphous materials, confined to experimental reports on carbon \cite{Shin}
or carbon and SiC fibers \cite{Gouadec}, without theory or consideration of dependence on strain pattern.
In previous work, Fabian and Allen \cite{Fabian} calculated the effect of hydrostatic pressure on the vibrational modes
of large supercells of a-Si (non-hydrogenated) via Stillinger-Weber classical potentials, but did not compute Raman spectra.
An \textit{ab initio} study \cite{Ribeiro} calculated vibrational modes (but not stress effects)
by density-functional theory,
but obtained Raman spectra only via semi-empirical bond polarizability models, which gave a significant discrepancy
from experiment.

Experimental work by Ishidate \textit{et al.} \cite{Ishidate} and Hishikawa \cite{Hishikawa} studied the effect of pressure and bending
on the Raman spectrum of a-Si:H.
However, it is not clear how to extract a strain coefficient (the general materials property) from these works,
due to insufficient detail about the experimental setups and stress applied \cite{Hishikawa_priv}.
Therefore only qualitative interpretations of a-Si:H stress from Raman spectroscopy have been possible \cite{Vetushka,KWu}.
\begin{figure*}
  \stackinset{l}{0.4 in}{t}{0.1 in}{\includegraphics[scale=0.15]{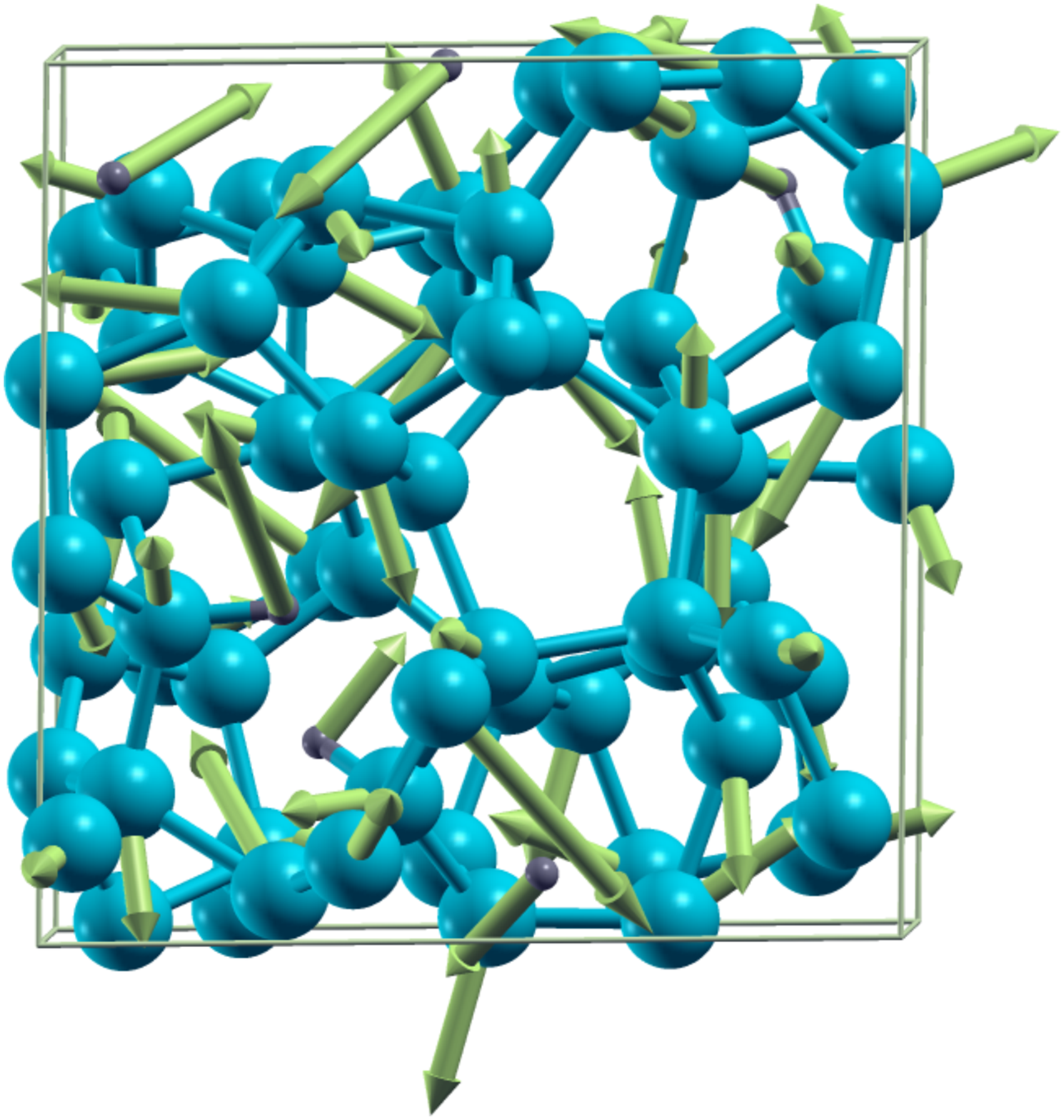}}{
    \stackinset{l}{2.0 in}{t}{0.9 in}{\includegraphics[scale=0.35]{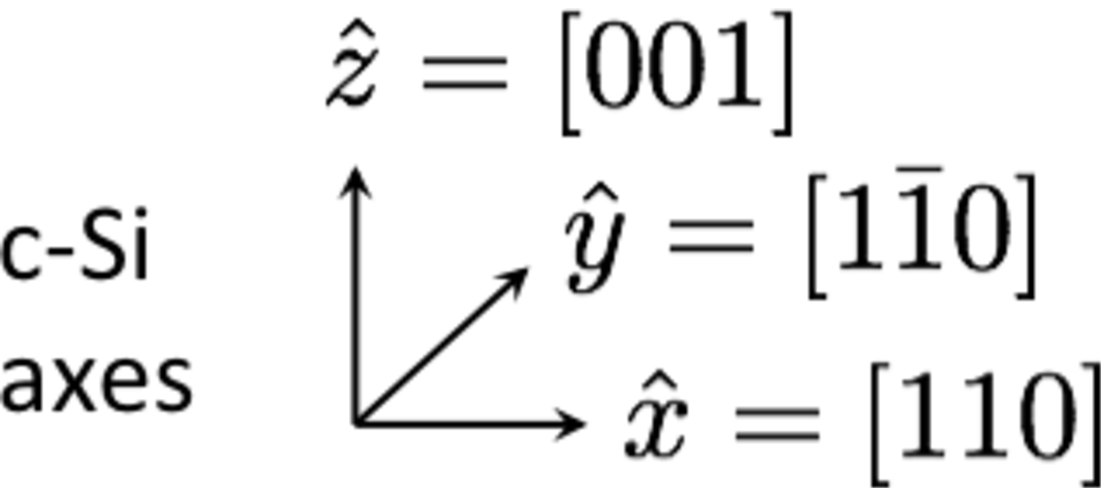}}{
      \stackinset{l}{1.7 in}{t}{0.1 in}{\includegraphics[scale=0.26]{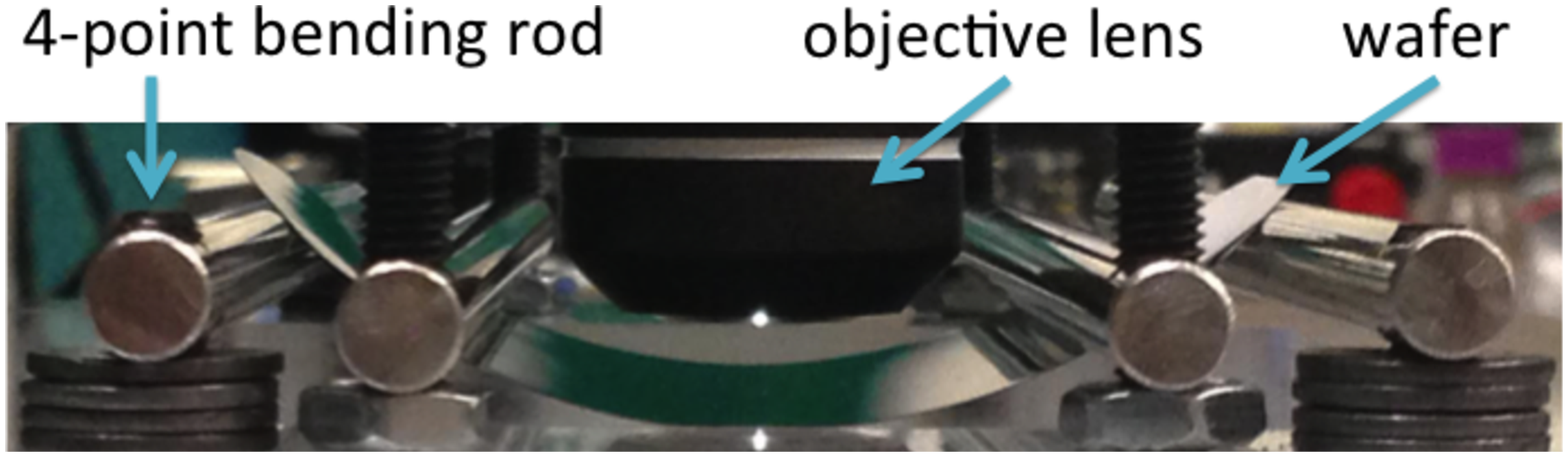}}{ %
    \includegraphics[scale=1.4]{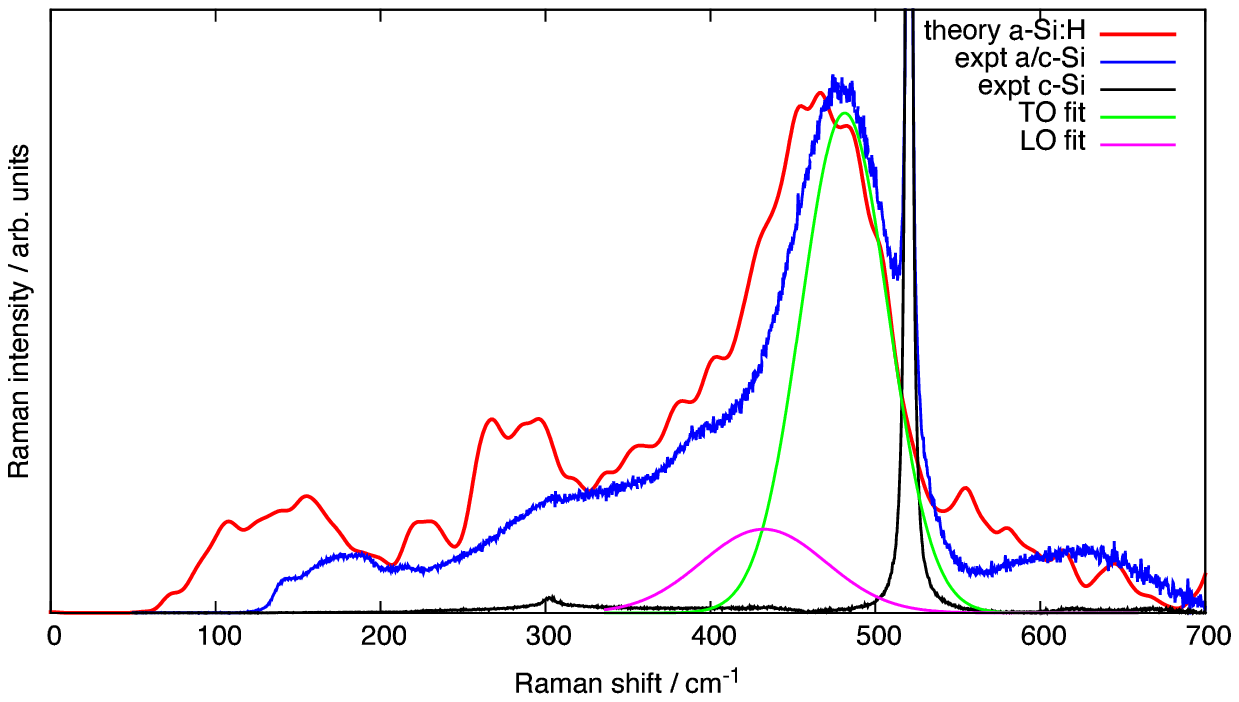}}}}
\caption{\label{fig:expt_neutral_Raman}
  Theoretically calculated Raman spectrum for a-Si:H, and measured Raman spectra for a-Si:H on c-Si, and c-Si,
  reduced by removal of the temperature- and frequency-dependent factors (see text),
  with fits to a-Si:H transverse and longitudinal optical peaks (TO, LO) in measured spectrum.
  Left inset: Example calculated Si$_{64}$H$_6$ TO vibrational mode. %
  Right inset: Experimental setup for Raman microscopy with four-point bending, and orientation of crystal axes in c-Si wafer.
}
\end{figure*}

In this Letter, we present a fully \textit{ab initio} computation of the Raman spectrum of a-Si:H under neutral and applied strain,
complemented with a systematic experimental study. 
We show the general form of peak shifts with strain in an amorphous material, and
obtain close agreement between theory and experiment in the spectra and the strain coefficient for the TO peak shift.
This provides the calibration needed for quantitative strain mapping of a-Si:H films for optical, electronic, and mechanical devices,
with sufficient sensitivity for applications of interest (analyzed in SI \cite{SI}).

Our theoretical calculations use an ensemble of periodic structures generated by the standard classical Monte Carlo Wooten-Winer-Weaire
approach \cite{WWW}, representing local regions which are averaged to find the overall properties of a-Si:H.
We add hydrogen to the sample by breaking randomly chosen Si-Si bonds at the beginning of the process, as in our previous work \cite{Johlin2013}
and implemented in our CHASSM code \cite{Strubbe}.
We use 34 structures to obtain a smooth Raman spectrum, each with formula Si$_{64}$H$_{6}$ to emulate a typical ~10\% hydrogen content, in a cube roughly 11 \AA\ on a side.
Density-functional theory (DFT) and density-functional perturbation theory (DFPT) \cite{Baroni}
calculations were performed with the Quantum ESPRESSO code (version 5.1) \cite{QE-2009} and the local-density approximation,
to obtain the phonons at ${\bf q} = \Gamma$ and their first-order Raman intensities \cite{Lazzeri}.
These widely used calculation methods have been found to be generally reliable for vibrational properties \cite{Baroni}.
Each structure was calculated also with $0.5$\% uniaxial compressive and tensile strain, which gave a resolvable effect within a linear regime. 
We study the unpolarized (isotropically averaged) Raman spectrum, with a Gaussian broadening of 5 cm$^{-1}$ standard deviation,
comparable to the separation between vibrational modes in an individual structure.

We benchmark the accuracy of our theoretical approach for strain effects on the Raman spectrum by calculations on c-Si under [100] uniaxial strain.
The Raman-active zone-center optical phonons have a frequency of 514 cm$^{-1}$, a typical DFT level of agreement with the experimental
value of 520 cm$^{-1}$ \cite{Parker}.
The slopes of the split modes are in reasonable agreement with the measured
values for bulk c-Si \cite{Anastassakis1990}, though slightly too small:
singly-degenerate, calculated -424 cm$^{-1}$ \textit{vs.} measured $p / 2 \omega^c_0 = -481 \pm 20\ {\rm cm}^{-1}$;
doubly-degenerate, calculated -547 cm$^{-1}$ \textit{vs.} measured $q / 2 \omega^c_0 = -601 \pm 20\ {\rm cm}^{-1}$.

For the experimental measurements, intrinsic a-Si:H films were deposited using a plasma-enhanced chemical vapor deposition tool
(PECVD, Surface Technology Systems) to a thickness of $\sim 1.1$ $\mu$m, on 3 inch diameter 100 $\mu$m ($\pm$ 15 $\mu$m) thick
$<$100$>$ c-Si wafers. Raman microscopy was performed using a Horiba LabRam-HR800 Raman spectrometer with a 632.8 nm excitation beam
focused to a 1 $\mu$m spot size. Compressive stress was applied to the a-Si:H film by bending the wafer in a custom-built four-point bending apparatus,
as shown in the inset of Fig. \ref{fig:expt_neutral_Raman}.

The obtained Raman spectra are shown in Fig. \ref{fig:expt_neutral_Raman}. The experimental results have been ``reduced'' by multiplication by the factor
$\omega \left( 1 - e^{-\hbar \omega / k T} \right)$ (where $\omega$ is the Raman shift and $T$ = 300 K is the temperature),
to be directly comparable to the calculated absolute Raman intensities \cite{Smith,Lazzeri},
in arbitrary units since we do not have an absolute intensity calibration.
The peaks in a-Si:H are conventionally named by the corresponding peaks in the vibrational density of states of c-Si \cite{Smith,KWu}.
The position of the transverse optical (TO) peak, the focus of this work, is at 470 cm$^{-1}$ (theory) and 480 cm$^{-1}$ (experiment), which agrees well within
the typical errors of DFT and the variation among a-Si:H samples \cite{Hishikawa}. An example calculated vibrational mode in the TO peak is shown in the inset
of Fig. \ref{fig:expt_neutral_Raman}. The longitudinal optical (LO) shoulder near 400 cm$^{-1}$ is also in good agreement.
The low-energy spectrum agrees less well due to the more delocalized modes \cite{Fabian} and sensitivity to the size of the calculated supercells.
The strong peak at 520 cm$^{-1}$ in the experiment is due to the underlying
c-Si substrate, which also has a small peak at 300 cm$^{-1}$ due to second-order
Raman scattering \cite{Temple}.
After a linear baseline correction, the experimental Raman spectra were fit to a sum of 3 Gaussians for the a-Si:H features \cite{Ishidate} and
a Lorentzian for c-Si \cite{Temple} according to standard practice; LO and TO fits shown in Fig. \ref{fig:expt_neutral_Raman} and SI \cite{SI}.
We underscore the significant improvement in theoretical agreement with experiment, compared to the previous DFT/semi-empirical Raman work \cite{Ribeiro} which
underestimated the TO peak by 50 cm$^{-1}$ and did not show the other peaks.
We can now quantitatively predict the strain effects on the spectrum.

\begin{figure}
  \includegraphics[scale=0.7]{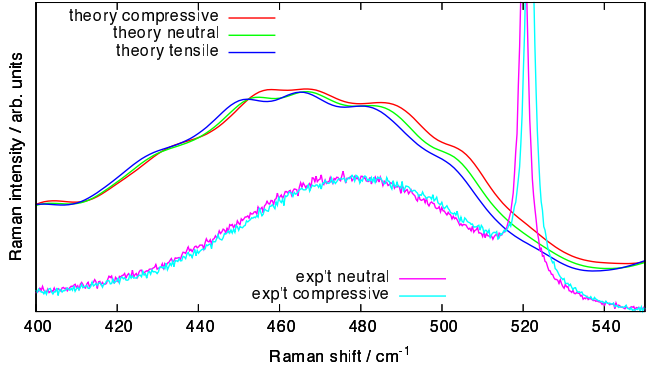} %
\caption{\label{fig:stressed_spectra}
  Effect of strain on the Raman spectra: theoretical calculations on a-Si:H with neutral strain and 0.5\% compressive and tensile strains,
  and a-Si:H on c-Si with neutral strain and 0.33\% compressive strain, with peaks blue-shifted by compressive strain and \textit{vice versa}.
}
\end{figure}

We now focus on the region $400 - 550\ {\rm cm}^{-1}$ around the a-Si:H TO peak and c-Si optical modes,
and add the calculated spectra under 0.5\% compressive and tensile uniaxial strains, and the measured spectrum under 0.33\% compressive uniaxial strain,
as shown in Fig. \ref{fig:stressed_spectra}.
The shifts to lower energies under tensile strain and higher energies under
compressive strain can be seen in theory and experiment, for both the a-Si:H and c-Si peaks.

To analyze the strain effect in a-Si:H in our calculation, we make a one-to-one correspondence between
the discrete vibrational modes in a supercell structure at each strain level.
We find the Raman intensity change with strain is
a small and almost uniform scaling over the spectrum. As a result, the strain effect on peak positions can be described by considering just the vibrational frequencies.
The $\pm 0.5$\% strain was confirmed to be in the linear regime by plotting frequencies over a range of strains.

For each mode in each structure, we compute the derivatives of the frequency in the compressive and tensile strain directions.
These derivatives are closely related to the mode Gr\"{u}neisen parameters $\gamma = - \frac{1}{\omega}\frac{d\omega}{d\epsilon}$,
and are shown in full in SI \cite{SI}.
We perform an average (weighted by the Raman intensities) over the derivatives of modes with frequencies
450 -- 490 cm$^{-1}$, yielding an overall TO peak position derivative of $460 \pm 10\ {\rm cm}^{-1}$.
The uncertainty is taken as the standard error of the mean,
taking only the different structures as independent.

It is difficult to determine the strain sufficiently accurately from our wafer curvature via Stoney's equation \cite{Freund},
and this would give only an averaged strain over the wafer.
Instead we use the c-Si Raman shifts as an internal calibration of the local strain at the beam spot. We exploit the fact that our Raman measurements
show both the a-Si:H thin film and the top of the underlying c-Si substrate (Fig. \ref{fig:expt_neutral_Raman}),
given the penetration depth of 1 ${\mathrm{\mu m}}$ for a-Si:H and c-Si \cite{Shah}.

To perform the calibration, we relate the
c-Si peak shift to uniaxial strain according to the approach of Refs. \onlinecite{Ganesan} and \onlinecite{DeWolf}.
The geometry of our four-point bending setup (inset in Fig. \ref{fig:expt_neutral_Raman}), results in uniaxial stress in [110] ($x$) in the roughly rectangular
region between the rods, according to the usual plane stress assumptions \cite{emech}.
The optical mode detected in our backscattering geometry is shifted from the unstrained frequency $\omega^c_0 = 520\ {\rm cm}^{-1}$ \cite{Parker}
proportionally to the strain $\epsilon_{xx}$ as $\Delta \omega^{c} = b \epsilon_{xx}$, where
\begin{align}
  b = \left[ -p \nu_{xz}^c + q \left( 1 - \nu_{xy}^c \right) \right] / 2 \omega^c_0 = -330 \pm 70\ {\rm cm}^{-1}
\end{align}
(more detail in SI \cite{SI}). We use c-Si Poisson ratios $\nu^c_{xy} = 0.064$ and $\nu^c_{xz} = 0.28$ \cite{Hopcroft},
and the strain coefficients $p = -1.25 \pm 0.25\ \left(\omega^c_0\right)^2$ and $q = -1.87\ \pm 0.37 \left(\omega^c_0\right)^2$ from an experiment with
the same 632.8 nm excitation as in this work \cite{Anastassakis1970}. Due to stress relaxation (\textit{i.e.} greater Poisson ratio $\nu^c_{zx}$) near the surface),
these strain coefficients are lower than those obtained at 1064 nm \cite{Anastassakis1990} with a signal penetrating about 100 $\mu$m into the bulk \cite{Shah}.

Next we connect the strain in c-Si to the strain in the a-Si:H film, specifically the trace ${\rm Tr}\ \epsilon^a$ (justified below).
Assuming no slip from the substrate, the strain $\epsilon_{xx}$ is the same in the a-Si:H film. Taking into account
the other directions,
\begin{align}
  {\rm Tr}\ \epsilon^a = d \epsilon_{xx} = \left( 1 - \nu_{xy}^{c} - \nu^{a} \right) \epsilon_{xx},
  \end{align}
where the coefficient $d = 0.69 \pm 0.05$, using $\nu^a = 0.25 \pm 0.05$ for dense films of 10\% H \cite{Kuschnereit}.

\begin{figure}
  \includegraphics[scale=0.68]{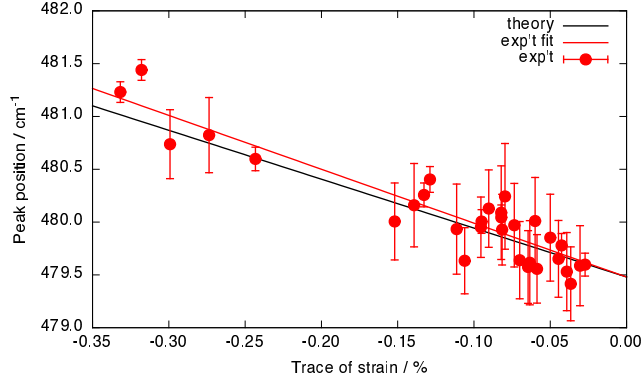} %
\caption{\label{fig:calibration}
  Shifts in a-Si:H Raman peak positions \textit{vs.} strain from uniaxial stress, inferred from c-Si peak shifts.
  Slopes: $-460 \pm 10\ \mathrm{cm}^{-1}$ (theory), $-510 \pm 120 \ \mathrm{cm}^{-1}$ (exp't fit). Both lines use experimental intercept.
  Relation can be used to infer local strain from Raman microscopy.
}
\end{figure}

We now infer strain for each position of the four-point bending setup from the c-Si peak shift
as ${\rm Tr}\ \epsilon^a = d \Delta \omega^c / b$.
Given a Young's modulus around 80 GPa \cite{Jiang} and strain 0.33\%, maximum stress was 260 MPa,
well within the range from PECVD growth \cite{Johlin2012}.
We plot the experimental a-Si:H peak position with respect to strain in Fig. \ref{fig:calibration}, showing a linear relationship
with regression slope $-510 \pm 120 \ \mathrm{cm}^{-1}$;
the uncertainty is mostly from the c-Si calibration values and $\nu^a$.
The plotted line with the theoretical slope (and experimental intercept) also fits the data well.
Note that if uniaxial strain rather than stress had been assumed in the wafer, we would have obtained
$b = q / 2 \omega^c_0$ and $d = 1$, yielding almost the same value $s = -520 \pm 110 \ \mathrm{cm}^{-1}$, showing
insensitivity to the exact mechanical boundary conditions.
We quote our result with respect to strain, rather than stress, to be more general since the shifts are due directly to bond length changes,
and the Young's modulus relating stress and strain can vary by a factor of 2 depending on synthesis conditions \cite{Jiang}.

Finally, we demonstrate the general form of the a-Si:H TO peak shift with strain. c-Si has a complicated dependence on the strain pattern due to its symmetry,
but a-Si:H is isotropic except at very short length scales.
For example, in our 70-atom cells, the calculated dielectric constant is $\sim 15$ with anisotropy only $\sim 0.6$.
Due to this effective symmetry, the calculated vibrational modes in the TO peak are delocalized, roughly isotropic,
and sensitive to Raman scattering in any polarization (see inset of Fig. \ref{fig:expt_neutral_Raman}).
Nonetheless, without any true symmetry, there is no counterpart to the three-fold degeneracy of the c-Si optical phonons.
As a result, the TO band transforms as a scalar rather than a vector as for c-Si.
Since there is no degeneracy there is no splitting as of the c-Si modes \cite{Anastassakis1970}.
In general, the frequency shift for such a scalar mode in a material would be $\Delta \omega = \sum_{ij} S_{ij} \epsilon_{ij}$
where $S$, like $\epsilon$, is a symmetric rank-2 tensor. For an isotropic material, symmetry dictates $S_{ij} = s \delta_{ij}$.
Therefore the peak shift is determined only by the trace of the strain tensor:
\begin{align} \label{eq-trace}
\Delta \omega^{a} = s \left( \epsilon_{xx}^{a} + \epsilon_{yy}^{a} + \epsilon_{zz}^{a} \right) = s\ {\rm Tr}\ \epsilon^a.
\end{align}
Indeed, we find in our calculations that the Raman spectrum is almost indistinguishable for applied uniaxial, biaxial, or triaxial strain tensors
with the same trace, even on a single 70-atom cell (see SI \cite{SI}). This analysis applies generally to isotropic amorphous vibrational modes.

\begin{table}
\begin{ruledtabular}
  \begin{tabular}{lllll}
                            & DFT & model & exp't       & classical \cite{Fabian} \\  \hline
    $\omega_c$ / cm$^{-1}$ & 514 &       & 520 \cite{Parker}     & 605 \\
    $\omega_a$ / cm$^{-1}$ & 470 &  430 \cite{Ribeiro} & 480       & 525 \\
    $\gamma_c$            & 0.98 &      & 1.08 \cite{Anastassakis1990} & 0.8 \\
    $\gamma_a$            & 0.98 &      & 1.06      & 1.0
\end{tabular}
\end{ruledtabular}
\caption{\label{table:comparison}
  Raman transverse optical (TO) peak positions $\omega$ and mode Gr\"{u}neisen parameters $\gamma$ for crystalline (c) and amorphous (a) Si,
  from: DFT, DFT plus bond polarizability model, experiment, and classical potentials. From this work unless cited.
}
\end{table}

We find that our theoretical ($-460 \pm 10\ \mathrm{cm}^{-1}$) and experimental ($-510 \pm 120\ \mathrm{cm}^{-1}$) values are consistent,
supporting the accuracy of the results. 
The agreement also implies lack of slip between the a-Si:H film
and c-Si substrate, as has been argued for thermal expansion of epitaxial graphene \cite{Ferralis},
as slip would relax strain and lower the measured coefficient.
The value is similar to the isotropic one for c-Si (surface), $-430 \pm 90\ {\rm cm}^{-1}$ \cite{Anastassakis1970}.
In Table \ref{table:comparison} we compare theoretical and experiment results
for peak frequencies and mode Gr\"{u}neisen parameters $\gamma$
of c-Si and a-Si:H. $\gamma$, describing anharmonicity, is important in the theory of thermal expansion and phonon transport \cite{Esfarjani}.
The importance of \textit{ab initio} calculations is shown by the much improved agreement with experimental $\omega$ and $\gamma$,
compared to classical potentials \cite{Fabian}.

To conclude, we obtained the Raman spectra of a-Si:H from first principles in good agreement with experiment. We computed the strain
coefficient for the TO peak from theory as $-460 \pm 10\ \mathrm{cm}^{-1}$, and measured a consistent value of $-510 \pm 120\ \mathrm{cm}^{-1}$,
achieving an experimental uncertainty similar to that for c-Si surfaces despite having to deconvolve much broader peaks.
We demonstrated, by symmetry analysis and explicit computation, the general form of strain effects on isotropic amorphous vibrational mode frequencies,
as $\Delta \omega = s\ {\rm Tr}\ \epsilon$, determined only by the trace of the strain.
The actual strain pattern (as in c-Si) needs to be provided by elasticity modeling \cite{DeWolf}.
Our results provide consistent and reliable calibration for the Raman/strain relation, enabling micro-Raman mapping of strain in a-Si:H films
for the further development of photovoltaic, electronic, and mechanical devices.

\begin{acknowledgments}

  We acknowledge Nicola Ferralis and Nouar Tabet for helpful discussions,
  and James Serdy for construction of the four-point bending setup.
  This work was supported by the Center for Clean Water and Energy at MIT and the King Fahd University of Petroleum and Minerals, Dhahran, Saudi Arabia
  under Project No. R1-CE-08.
Computation was performed at the National Energy Research
Scientific Computing Center at Lawrence Berkeley National Laboratory, a DOE Office of Science User Facility 
supported by the Office of Science of the U.S. Department of Energy 
under Contract No. DE-AC02-05CH11231.
Fabrication was performed at the Center for Nanoscale Systems (CNS) at Harvard University,
a member of the National Nanotechnology Infrastructure Network (NNIN), supported by the National Science Foundation under NSF award no. ECS-0335765.

\end{acknowledgments}

%
%
%


\pagebreak
\widetext
\begin{center}
\textbf{\large Supplementary Information}
\end{center}
\setcounter{equation}{0}
\setcounter{figure}{0}
\setcounter{table}{0}
\setcounter{page}{1}
\makeatletter
\renewcommand{\theequation}{S\arabic{equation}}
\renewcommand{\thefigure}{S\arabic{figure}}

\section{Theoretical calculations}

\begin{figure*}[b]
  \includegraphics[scale=1.2]{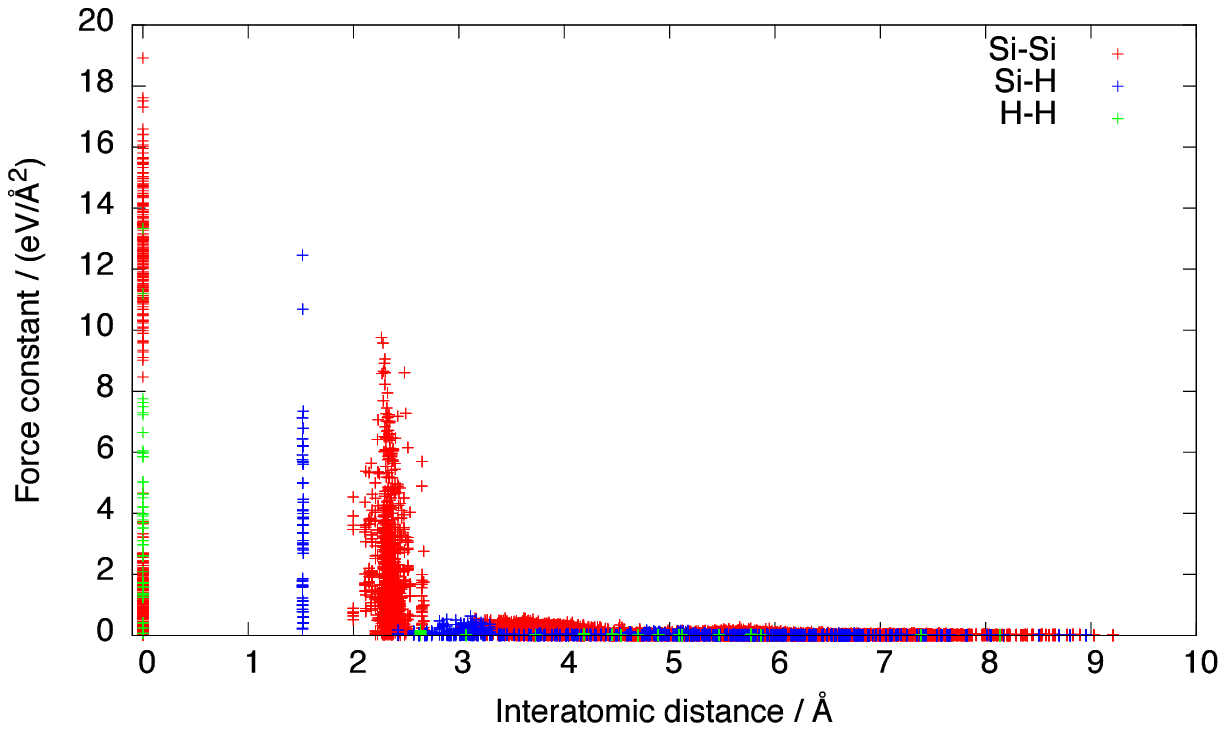}
\caption{\label{fig:matdyn_r}
  Force constants (dynamical matrix elements) from vibrational calculation, as a function of the interatomic distance.
}
\end{figure*}

Density-functional theory (DFT) calculations were performed with the local-density approximation (LDA)
in the Perdew-Zunger parametrization \cite{PZ}, and norm-conserving pseudopotentials.
The plane-wave cutoff was 60 Ry and a $2 \times 2 \times 2$ Monkhorst-Pack $k$-grid was used. Beginning with the
WWW structures, the cell parameters and atomic positions are optimized with a variable-cell relaxation until forces were less
than $10^{-4}$ atomic units and components of the stress tensor were less than 0.1 kbar.
We relaxed the atomic coordinates of the strained structures with fixed cell parameters, to take into account the important atomic rearrangements
under strain \cite{Fabian}. The atomic masses used were 28.0855 amu for Si and 1.00794 amu for H.
The total of 34 structures was
found to be sufficient to give a converged and fairly smooth average Raman spectrum. We use only 64 Si atoms due to the computationally intensive nature
of phonon calculations. Larger structures (e.g. 216 Si atoms) have been possible when calculating only the ground state and when seeking longer-ranged
properties such as the deepest hole trap \cite{Wagner,Johlin2013}, but this smaller cell is adequate for the vibrational spectrum (as shown in
previous work \cite{Ribeiro}).
The dynamical matrix as a function of interatomic distance for one structure is plotted in Fig. \ref{fig:matdyn_r},
showing that the vibrational interactions are short-ranged,
being quite small beyond the nearest neighbors, and are therefore well accounted for in a cell of 11 \AA\ on a side.

For the c-Si benchmark calculation, a $5\times5\times5$ {\bf k}-grid and an LDA-optimized
lattice parameter of 5.380 \AA\ were used.  Applying uniaxial strains up to $\pm 1\%$, compressive and tensile, in the [100] direction, we found modes
varying linearly and splitting into a singlet and doublet as the symmetry is broken.

\begin{figure*}
  \includegraphics[scale=1.2]{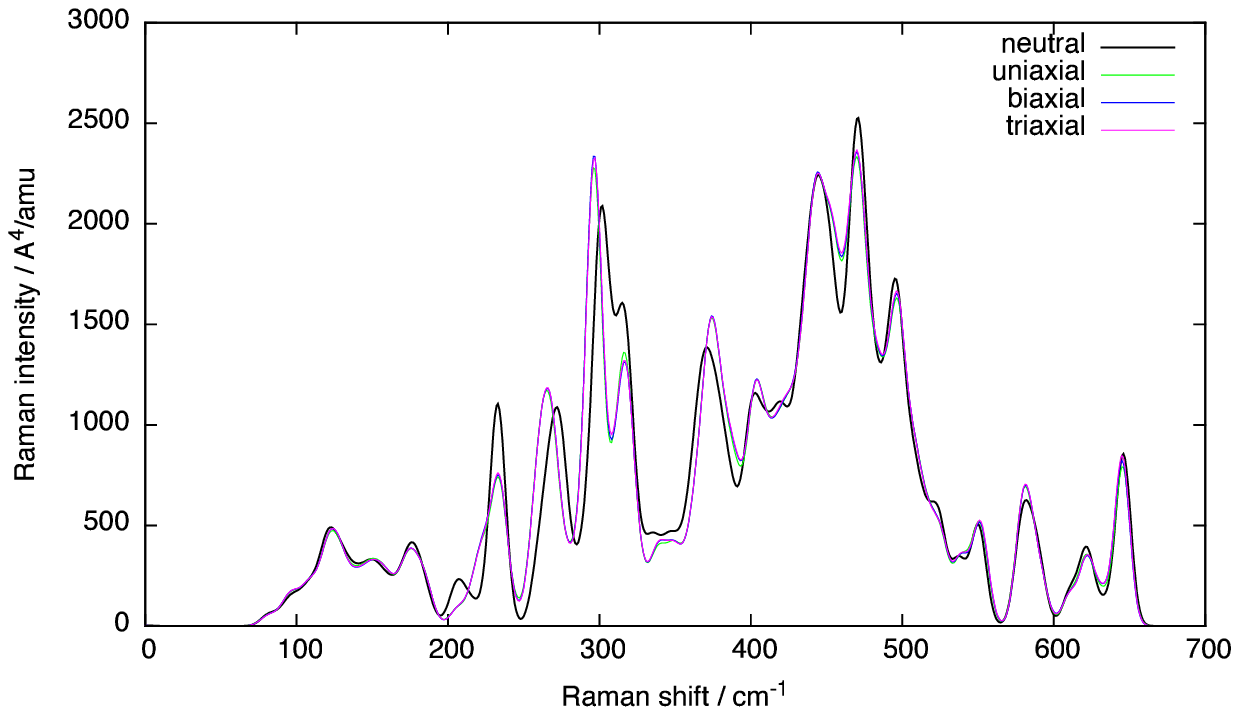}
\caption{\label{fig:stress_patterns}
  Calculated Raman spectra of one 70-atom structure under neutral strain, and with compressive uniaxial, biaxial, and triaxial strain
(each with the same trace, -0.2\%), showing the dependence only on the trace of the strain.
}
\end{figure*}

The calculated Raman spectra of one 70-atom structure under neutral strain, and with compressive uniaxial, biaxial, and triaxial strain
(each with the same trace ${\rm Tr}\ \epsilon$ = -0.2\%) are shown in Fig. \ref{fig:stress_patterns}.
There is remarkably little difference between the spectra under different strain patterns, despite a clearly visible difference from
the neutral-strain spectrum. This demonstrates the validity of our conclusion
from symmetry analysis that the peak shifts are affected only by the trace of the strain for delocalized amorphous vibrational modes, even
within a single small cell.

\begin{figure*}
  \includegraphics[scale=0.65]{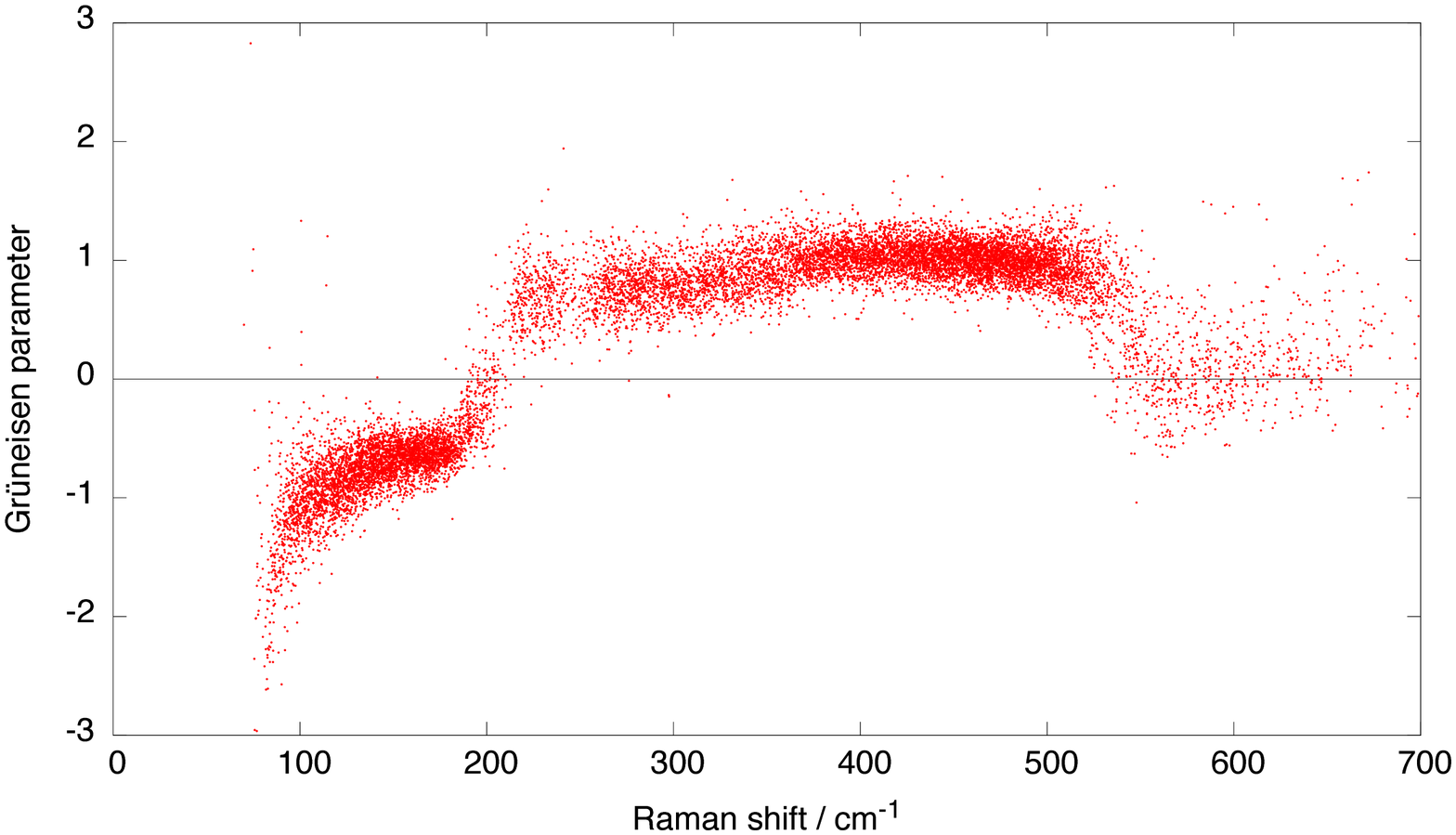}
\caption{\label{fig:gruneisen}
Calculated mode Gr\"uneisen parameters of vibrational modes in all structures.
}
\end{figure*}

The calculated mode Gr\"uneisen parameters for all 210 vibrational modes (up to 700 cm$^{-1}$) of each of the 34 structures and two directions of strain are shown
in Fig. \ref{fig:gruneisen}. The values are around
1 for the LO and LA peaks (in agreement with the classical-potentials study \cite{Fabian}). Our simulations found
values around $-1$ for the TA peak, in contrast to values around 0 in the classical-potentials study.
We find that the Si-H modes above 700 cm$^{-1}$ have Gr\"uneisen parameters around 0, since they are localized
and not particularly affected by strain.

\section{Experimental methods}

Plasma-enhanced chemical vapor deposition was done at 200 C using pure silane gas at a flow rate of 55 sccm, a discharge power of 30 W at a frequency of 13.56 MHz,
and a chamber pressure of 400 mTorr to minimize the intrinsic stress in the films \cite{Johlin2012}.
A 632.8 nm excitation beam was attenuated using a
neutral density filter to a power of $\sim$0.6 mW, to prevent crystallization of the a-Si:H films. Initial traces were taken
with lower transmittance to ensure no partial crystallization at the measurement fluence was occurring. After attenuation, the beam was
focused to approximately a 1 $\mu$m diameter spot size using a 100$\times$ long-working-distance objective. Traces were averaged over
10 acquisitions each, with a 5 second integration time to reduce the noise in the spectra, and a resolution of $\sim$0.3 cm$^{-1}$.
Traces were taken from three separate runs on the same wafer, with 14, 9, and 6 different stress states for the respective runs,
with the last run beginning at a pre-loaded condition and driven until mechanical failure of the substrate.

\begin{figure*}
  \includegraphics[scale=1.2]{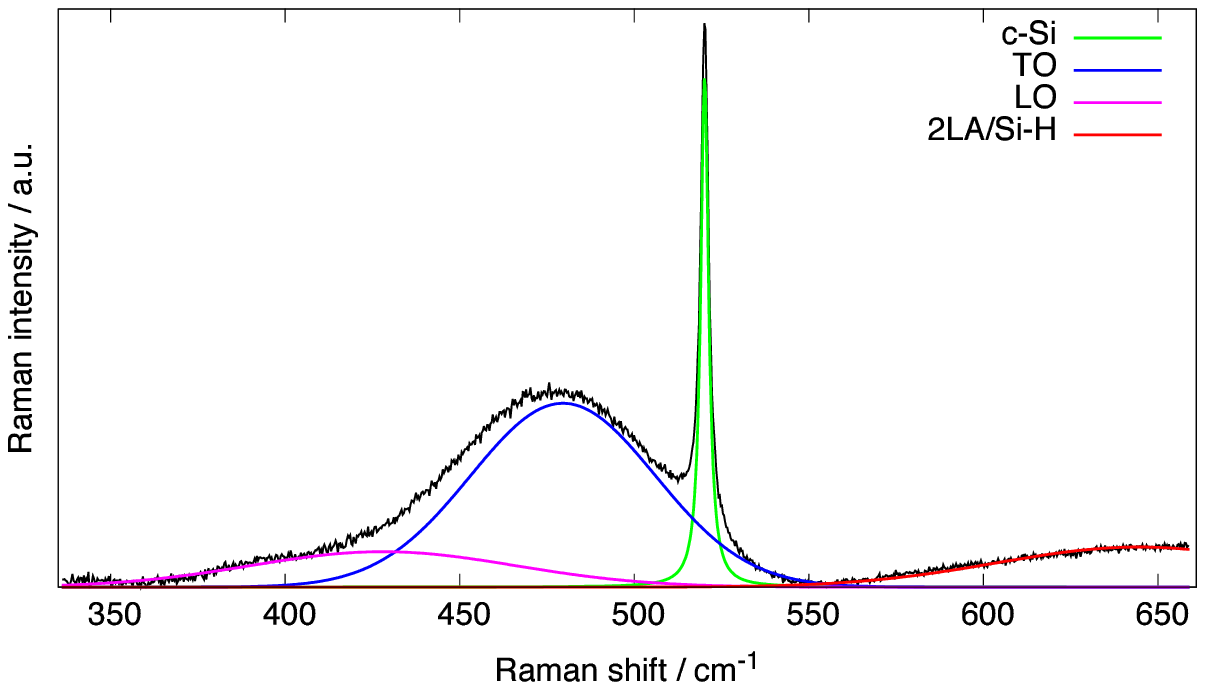}
\caption{\label{fig:deconv}
  Deconvolution of experimental Raman spectrum (after linear baseline correction) into Gaussians for the transverse
  optical (TO), longitudinal optical (LO), and second-order longitudinal acoustic (2LA)/Si-H peaks of a-Si:H,
  and a Lorentzian for the underlying c-Si substrate's optical phonons.
}
\end{figure*}

From the fitting procedure (Fig. \ref{fig:deconv}), we obtain uncertainties for the TO peak position typically $\sim$0.3 cm$^{-1}$, while
those for the sharper c-Si peak are about 0.005 cm$^{-1}$ and negligible for the analysis.
The contributions near $550-700\ \mathrm{cm}^{-1}$ (also seen in the calculation) may be
assigned to Si-H bonds and second-order scattering from LA (2LA) \cite{Brodsky} (though only the first effect is accounted for in the calculation),
and perhaps further background.

\section{Analysis of strain and c-Si peak shift}

The wafer has the standard circular form with a flat along the [1$\bar{1}$0] direction.
We oriented the wafer so that uniaxial stress was applied along the [110] direction ($x$), and the wafer surface normal was [001] ($z$).
The four-point bending rods lay along the [1$\bar{1}$0] direction ($y$), as shown in the inset of main text Fig. 1.

Given the negligible stress in the $y$ and $z$ directions, the strain in these
directions in the c-Si substrate is related to that in the $x$-direction by the Poisson ratios:
\begin{align}
  \left\{ \begin{array}{l}
    \epsilon_{yy}^{c} = -\nu_{xy}^{c} \epsilon_{xx}^{c} \\
    \epsilon_{zz}^{c} = -\nu_{xz}^{c} \epsilon_{xx}^{c}
    \end{array} \right.
\end{align}
Due to the small a-Si:H film thickness compared to the wafer thickness, and continuity of the strain parallel to the interface (assuming no slip),
$\epsilon_{xx}$ and $\epsilon_{yy}$ at the top of the c-Si substrate are equal to those in the a-Si:H thin film.
\begin{align}
  \left\{ \begin{array}{l}
  \epsilon_{xx}^{a} = \epsilon_{xx}^{c} \equiv \epsilon_{xx} \\
  \epsilon_{yy}^{a} = \epsilon_{yy}^{c} \equiv \epsilon_{yy}
  \end{array} \right.
\end{align}

Since stress normal to the interface is continuous (and zero in this case), the strain normal to the interface is not continuous.
The value in a-Si:H is related to its (isotropic) Poisson ratio: $\epsilon_{zz}^{a} = -\nu^{a} \epsilon_{xx}$.
The shear strains in this coordinate system ($\epsilon_{xy}, \epsilon_{xz}, \epsilon_{yz}$) are negligible in each material.

We analyze the expected peak shift in c-Si according to the approach of Refs. \onlinecite{Ganesan} and \onlinecite{DeWolf}.
There are three optical phonon modes in c-Si, degenerate at $q=\Gamma$ in the absence of strain.
According to group theory, these modes belong to representations transforming as $R_x$, $R_y$, and $R_z$.
Backscattering from the (001) surface is only sensitive to the $R_z$ mode,
and there is no mixing between the $R_z$ mode and the other two in the absence of shear involving the $z$ direction,
so we do not observe strain-induced splitting of the c-Si peak \cite{Anastassakis1970}.
The eigenvalue of the secular equation for the $R_z$ mode is
\begin{align}
  \lambda = p \epsilon_{33}^c + q \left( \epsilon_{11}^c + \epsilon_{22}^c \right) = p \epsilon_{zz}^c + q \left( \epsilon_{xx} + \epsilon_{yy} \right) \\ \nonumber
  = \left[ -p \nu_{xz}^c + q \left( 1 - \nu_{xy}^c \right) \right] \epsilon_{xx}
\end{align}
The peak shift from $\omega^c_0$, the frequency at neutral strain, is given by
\begin{align}
  \Delta \omega^{c} = \omega^c - \omega^c_0 = \lambda / 2 \omega^c_0 = b \epsilon_{xx}
\end{align}

For the strain coefficients, there are several values available in the literature from uniaxial stress experiments,
based on different excitation wavelengths. The experiment
at 632.8 nm \cite{Anastassakis1970} obtained $p = -1.25 \pm 0.25\ \left(\omega^c_0\right)^2$ and $q = -1.87\ \pm 0.37 \left(\omega^c_0\right)^2$,
and was sensitive only to the
near-surface region. An experiment at 1064 nm from some of the same authors obtained \cite{Anastassakis1990} instead
$p = -1.85 \pm 0.06\ \left(\omega^c_0\right)^2$ and $q = -2.31\ \pm 0.06 \left(\omega^c_0\right)^2$, more precise and somewhat higher.
The penetration depth is about 100 $\mu$m at 1064 nm rather than 1 $\mu$m at 632 nm \cite{Shah}, so the signal came from the bulk;
they argue that at the surface, there was stress relaxation, giving lower values.
In this work, since we are using 632.8 nm and only measuring the near-surface region, we calibrate using the experiment
with those conditions \cite{Anastassakis1970}, since the same out-of-plane surface stress relaxation in c-Si
(\textit{i.e.} greater local Poisson ratio $\nu^c_{xz}$) should be occurring in our unaxially stressed wafer.
Since a-Si:H is prepared in thin-film form and not as a bulk material, this $\sim 1 \mu$m near-surface region is the only region of interest.

\section{Practicability of strain mapping}

In this section we analyze the practicability of strain mapping of a-Si:H films by Raman microscopy using the coefficient determined in this work,
showing that the situation is not very different from c-Si for which there are many successful applications of the technique \cite{DeWolf}.

To find an absolute level of strain, one needs a reference unstrained peak position. In the case of c-Si, reported values in the literature are
$519 \pm 1\ \mathrm{cm}^{-1}$ \cite{Temple} and $520 \pm 0.5\ \mathrm{cm}^{-1}$ \cite{Parker}. The uncertainties in this reference position are much larger
for a-Si:H -- the TO peak is in the range $470-480\ \mathrm{cm}^{-1}$, varying systematically and reproducibly
due to deposition conditions such as temperature, substrate, and hydrogen
content \cite{Hishikawa}. Na\"{i}ve use of such literature values as a reference would lead to a corresponding large uncertainty in the inferred absolute strain.
Typically absolute levels of strain are not the quantity of interest in a Raman microscopy study, however. The point is to show
spatial variation in strain, and changes due to heating, etching, deposition, crystallization, etc., for which an absolute reference is not required. For
example, c-Si strain mapping is commonly used in situations where the peak shifts of interest are only 0.1 cm$^{-1}$, much less than the uncertainties in
the reference position, because only relative amounts of strain are being studied \cite{DeWolf}. If an absolute strain measurement is really needed,
the uncertainty in the reference peak position can be greatly reduced by performing a control
experiment in which a film is deposited under the same conditions, any strain is relieved by etching away the substrate, and the
Raman peak position is measured for the unstrained film \cite{Shin} to find the reference. Another approach is to measure the average strain in the film
via substrate curvature and Stoney's equation (the typical strain measurement on thin films \cite{Freund}), or via X-ray diffraction \cite{XLWu},
and then use Raman microscopy to measure the local deviations from the average.

Peak shifts can be mapped to strain with a single coefficient over a certain linear regime. Our experimental measurements had a strain range of
0.33\% (equivalent to 1.7 cm$^{-1}$ or 260 MPa), and our set of theoretical calculations used $\pm 0.5$\% ($\pm$2.3 cm$^{-1}$ or $\pm$400 MPa), showing linearity
over this range. These ranges were chosen as the maximum before fracture of the c-Si substrate in our 4-point bending setup, and a value close to levels of experimental
interest that gave sufficiently resolvable peak shifts in the calculations.
In fact, our further theoretical calculations showed the linear regime extends to at least $\pm 1$\% strain ($\pm$4.6 cm$^{-1}$ or $\pm$800 MPa).
Therefore strain mapping can be applied straightforwardly throughout this range. At some yet higher strain level, the higher-order anharmonicity of the interatomic
potential and the atomic rearrangements under strain may cause a nonlinear strain-frequency relation, and the Raman intensities themselves may be altered.
Strain mapping could be still be possible in this regime with a higher-order polynomial model.

The sensitivity of strain mapping is limited by the uncertainty in the determination of the Raman peak position, setting the smallest strain differences that
can be resolved. We had an error of 0.3 cm$^{-1}$ for deconvolution
of the broad Gaussian peak of a-Si:H. This should be compared to 0.05 cm$^{-1}$, the best resolution obtainable for c-Si peak shifts according to De Wolf \cite{DeWolf}.
That resolution is limited not only by uncertainty in peak fitting (we had a much smaller error of 0.005 cm$^{-1}$ for the sharp Lorentzian of c-Si), but also the stability
over time of the Raman spectrometer, which should apply equally to measuring a-Si:H in its nearby frequency range. Thus a-Si:H uncertainty is only 6 times less than
the quoted best obtainable for c-Si as of 1996, a level that has proved to be sufficient for a wide range of successful applications of Raman strain mapping.
In fact, Raman peak shifts in a-Si:H over ranges much larger than this level are commonly reported: 9 cm$^{-1}$ from thermal stress \cite{KWu}, 2.5 cm$^{-1}$
from cantilever bending \cite{Vetushka}, 8 cm$^{-1}$ from laser heating for crystallization (the original motivation for this work) \cite{Tabet}.
These shifts imply stress up to 1.5 GPa. A sensitivity of 0.3 cm$^{-1}$ (about $6 \times 10^{-4}$ strain or 50 MPa) is more than sufficient to infer strain differences
from the Raman spectra measured in these works. Moreover, while our uncertainty
was sufficiently small that it did not dominate the ultimate uncertainty of the measured Raman strain coefficient, it is not
necessarily the best that can be achieved if one sets out to resolve small strains. There may be scope for improvement with better Raman spectrometer resolution
and better signal-to-noise ratio from longer integration times or increased intensity (limited by the need to avoid heating the a-Si:H to the point of crystallization).

In c-Si, the actual strain tensor is generally accessible only through elasticity modeling \cite{DeWolf} (or to some degree from Raman measurements with
high numerical aperture and polarizers \cite{Bonera}). A given peak shift, due to the splitting of the three degenerate optical phonons, could be caused by
various different strain magnitudes and patterns, as determined by solving a quadratic equation for the eigenvalues \cite{Ganesan}.
One can only determine what the strain magnitude is by assuming strain patterns such as uniaxial, biaxial, or triaxial with respect to particular axes,
without any simple way of summarizing this information \cite{DeWolf}. We have the same issue for a-Si:H
that the strain pattern is left undetermined by the peak shift measurement. However, since there is no degeneracy and the shift is just proportional to the
trace as $\Delta \omega = s\ {\rm Tr}\ \epsilon$, we can determine the trace (a mathematical invariant) of the stress tensor, and express the conclusion in closed form.
Thus a-Si:H strain mapping has the advantage of a simpler analysis.

Raman microscopy has many advantages as a strain measurement method, even beyond the unique ability to obtain spatial resolution.
It is non-destructive, unlike methods involving etching away a substrate \cite{Shin};
it uses a technique already widely employed on a-Si:H for other purposes \cite{Mahan,KWu,Tabet,Hishikawa};
it can be used to study the material at varying depths \cite{Paillard} depending on the chosen excitation
energy \cite{Anastassakis1990}, including when buried under another sufficiently transparent material;
it can simultaneously measure the strain in different materials in the device, such as c-Si in the substrate as we have done here;
it can be used to extract local temperatures too via the Stokes/anti-Stokes intensity ratio \cite{Gu},
and it can be used to monitor strain in real time \cite{Pomeroy}.

Given these considerations, we believe our results open the way to many useful applications of Raman microscopy for strain mapping
in a-Si:H, and indeed extension to other amorphous materials.

\end{document}